\documentclass[
    aps,
    reprint,
    superscriptaddress,
    amsmath,
    amssymb,
    amsthm,
    floatfix,
    onecolumn,
    natbib
    ]{revtex4-2} 
\usepackage{graphicx} 
\usepackage{hyperref}
\usepackage{adjustbox}
\usepackage{array}
\usepackage{amsfonts}

\usepackage{xcolor}

\DeclareMathOperator*{\argmin}{argmin}
\DeclareMathOperator*{\argmax}{argmax}
\newtheorem{theorem}{Theorem}

\begin{document}

\title{Optical Quantum Sensing for Agnostic Environments via Deep Learning}

\author{Zeqiao Zhou}
\affiliation{Department of Electronic Engineering and Information Science, University of Science and Technology of China, Hefei 230027, China}

\author{Yuxuan Du}
\affiliation{JD Explore Academy, Beijing 101111, China}

\author{Xu-Fei Yin}
\affiliation{Hefei National Research Center for Physical Sciences at the Microscale and School of Physical Sciences, University of Science and Technology of China, Hefei 230026, China}
\affiliation{CAS Center for Excellence in Quantum Information and Quantum Physics, University of Science and Technology of China, Shanghai 201315, China}

\author{Shanshan Zhao}
\affiliation{JD Explore Academy, Beijing 101111, China}

\author{Xinmei Tian}
\affiliation{Department of Electronic Engineering and Information Science, University of Science and Technology of China, Hefei 230027, China}

\author{Dacheng Tao}
\affiliation{School of Computer Science, The University of Sydney, NSW 2008, Australia}

\begin{abstract}
Optical quantum sensing promises measurement precision beyond classical sensors termed the Heisenberg limit (HL). However, conventional methodologies often rely on prior knowledge of the target system to achieve HL, presenting challenges in practical applications. Addressing this limitation, we introduce an innovative Deep Learning-based Quantum Sensing scheme (DQS), enabling optical quantum sensors to attain HL in agnostic environments. DQS incorporates two essential components: a Graph Neural Network (GNN) predictor and a trigonometric interpolation algorithm. Operating within a data-driven paradigm, DQS utilizes the GNN predictor, trained on offline data, to unveil the intrinsic relationships between the optical setups employed in preparing the probe state and the resulting quantum Fisher information (QFI) after interaction with the agnostic environment. This distilled knowledge facilitates the identification of optimal optical setups associated with maximal QFI. Subsequently, DQS employs a trigonometric interpolation algorithm to recover the unknown parameter estimates for the identified optical setups. Extensive experiments are conducted to investigate the performance of DQS under different settings up to eight photons. Our findings not only offer a new lens through which to accelerate optical quantum sensing tasks but also catalyze future research integrating deep learning and quantum mechanics.  
\end{abstract}

\maketitle

\section{Introduction}

Quantum sensing, by measuring physical quantities with a precision approaching Heisenberg limit (HL), has emerged as a leading practical application of quantum technology \cite{giovannetti2011advances,pirandola2018advances}, which offers new opportunities in various fields, including medicine, navigation, optical time transfer, and energy detection \cite{aslam2023quantum,Feng_2019, caldwell2023quantum,crawford2021quantum}. Achieving the HL demands a quantum sensing scheme meticulously tailored to the target quantity, encompassing optimal probe preparation and parameter estimation \cite{polino2020photonic,barbieri2022optical}. Linear optics, given its robustness against noise and decoherence combined with its long-range communication capability, stands out as a premier platform for quantum sensing \cite{qin2023unconditional}. Nevertheless, in various practical optical scenarios characterized by uncontrolled or unknown quantum systems such as biological processes or mineral and oil detection, formulating the optimal scheme becomes a formidable challenge \cite{thomas2011real,bongs2019taking,xavier2021quantum}. Despite the promising enhanced precision offered by optical quantum sensing, it remains uncharted territory to design the optimal scheme for such agnostic settings.

Deep learning has recently been demonstrated as a powerful tool to learn quantum systems with partial or incomplete information \cite{gebhart2023learning}. Initial works have focused on efficiently learning quantum systems from local measurement results, reconstructing the quantum state with various architectures of neural networks \cite{torlai2018neural,carrasquilla2019reconstructing,palmieri2020experimental,ahmed2021quantum}. These efforts further extend to the quantification of essential properties in quantum systems, such as entanglement \cite{gao2018experimental,yin2022efficient,koutny2023deep} or fidelity \cite{zhang2021direct,wu2023quantum,qian2023multimodal}. Concurrently, other works have focused on learning interpretable representations for quantum systems \cite{zhu2022flexible} or quantum experiments \cite{flam2022learning, jaouni2023deep}, discovering novel insights from unlabeled data via a data-driven paradigm. With the continuous advancement of both algorithmic frameworks and hardware infrastructures, deep learning holds the appealing potential to facilitate optical quantum sensing in realistic applications.

\begin{figure*}[!t]
\centering
\includegraphics[width=0.98\textwidth]{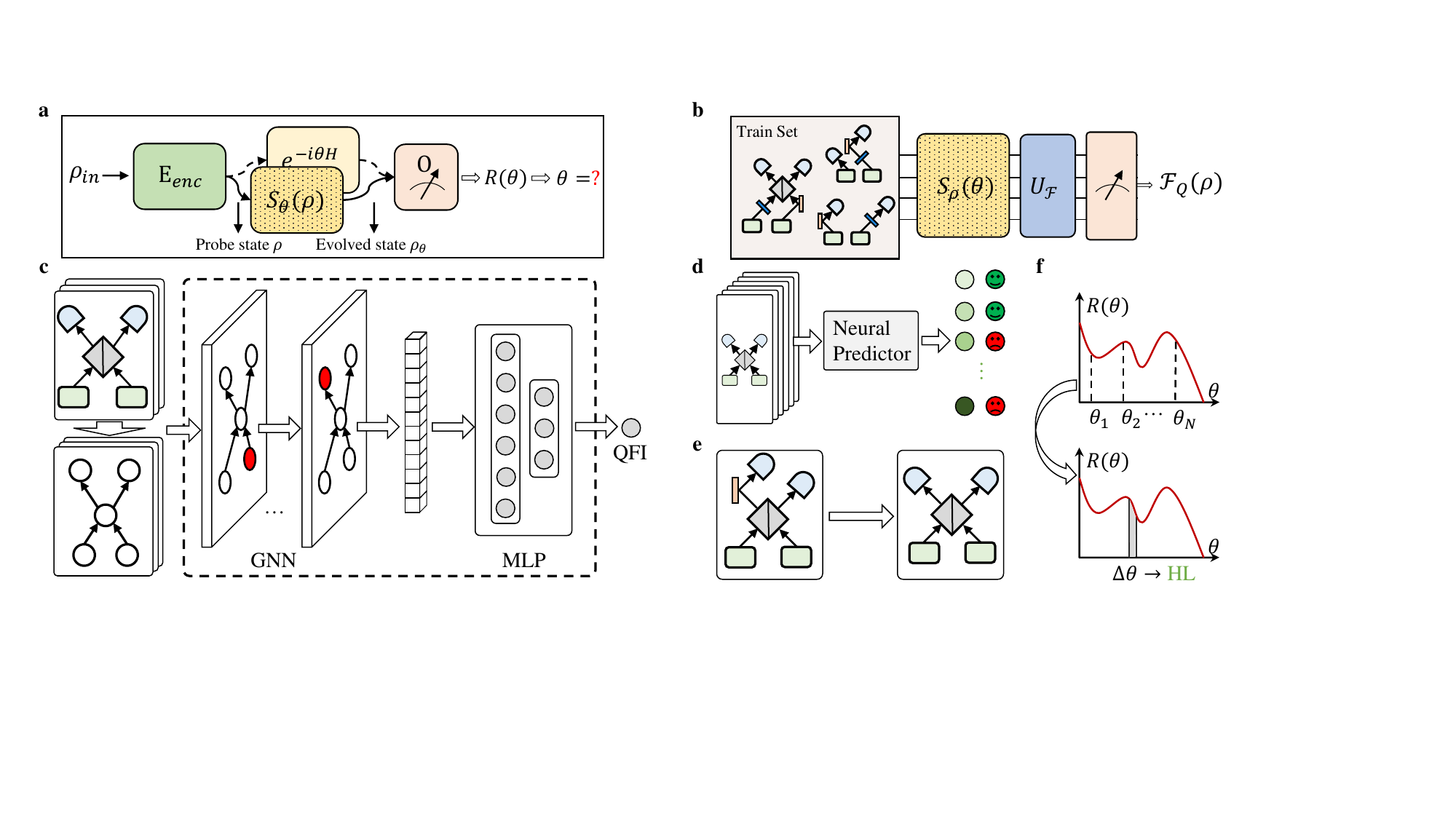}
\caption{\textbf{Schematic of deep learning based quantum sensing.} \textbf{a. Optical quantum sensing.} The probe state $\rho$, prepared by an optical setup, is sent through a channel that encodes parameter $\theta$ and then measurement operator $O$, resulting in a response function $R(\theta)$. The ``black-box'' channel is depicted with a dotted yellow box such that the explicit form of $H$ is unknown. This contrasts with the scenario of the informed environment in which the expression of $H$ is known, highlighted by the light yellow box. \textbf{b. Data collection.} Quantum fisher information of a probe state is obtained via queries to an oracle $\mathcal{U}_\mathcal{F}$. \textbf{c. Training.} A graph neural network is trained to predict the quantum fisher information on labeled data. \textbf{d. Ranking.} Randomly sample massive unlabeled setups and use pretrained GNN to rank the QFI of them. \textbf{e. Fine-tuning.} Fine-tune the selected setups, such as the layout and devices, and remove unnecessary ones. \textbf{f. Response Inferring.} Use trigonometric interpolation to infer the response function $R(\theta)$ and analyze the uncertainty $\Delta \theta$ of it.}
\label{fig:1}
\end{figure*}

In this work, we propose an end-to-end, Deep learning-based Quantum Sensing (DQS) scheme for agnostic environments. Striving for the utmost precision limit, DQS navigates the vast space of optical probe states to identify the optimal configuration by maximizing the Quantum Fisher Information (QFI) of the state after interacting with the environment, an intrinsic determinant of sensing precision. To achieve this, DQS employs a Graph Neural Network (GNN) to learn the mapping rule between the optical configuration and the corresponding QFI using few offline data. The trained GNN efficiently searches for the optical configuration with the highest QFI among exponentially numerous candidates. Once the optimal optic setup for the probe state is identified, DQS integrates a post-processing method to estimate the target parameter from the measurement results of the probe state in agnostic settings. In numerical experiments, through evaluating the DQS on datasets with different sizes and numbers of photons, we observe that DQS can perform well with a relatively small amount of training examples, and even discover the optimal probes when not encountering them during training. Furthermore, DQS saturates HL in an eight-photon quantum sensing task, while a trivially designed scheme fails. Our proposal can be readily adapted to other relevant optical tasks, opening up exciting prospects for future research on quantum technology applications in realistic settings.

\section{Deep learning-based Quantum Sensing scheme}

Here we consider single-parameter linear optical quantum sensing in real-world scenarios. As shown in Figure \ref{fig:1}a, the process begins with an initial state $\rho_{in}$. The implementation of an optical setup acts as an encoding channel $\mathrm{E}_{enc}$ to prepare an $N$-qubit probe state $\rho= \mathrm{E}_{enc}(\rho_{in})$ where $\rho_{in},\rho \in \mathbb{C}^{2^N \times 2^N}$. Subsequently, this probe state interacts with the environment via a channel given by $ S_\theta(\rho) = e^{-\frac{1}{2}i\theta H}\rho e^{\frac{1}{2}i\theta H} $ where $\theta$ represents the parameter of interest, such as a magnetic field, and $H$ is the encoding Hamiltonian associated with it. Following this, the evolved state, $\rho_{\theta} = S_\theta(\rho)$, is measured against an observable operator $O$, yielding a \textit{response function} expressed as 
\begin{equation}
 R(\theta) = \mathrm{Tr}(O\rho_{\theta}).
\end{equation} 
The primary objective here is to estimate $\theta$ with the highest precision from $R(\theta)$, thereby minimizing the uncertainty $(\Delta \theta)^2$. Under an optimal estimation procedure, this uncertainty is bounded by $ (\Delta \theta)^2 \sim  1/\mathcal{F}_Q(\rho_\theta)$ with $\mathcal{F}_Q(\rho_\theta)$ representing the quantum Fisher information (QFI) of $\rho_\theta$ with respect to $\theta$ \cite{rao1992information}. For a probe state with classical correlations, the QFI scales linearly with $N$, leading to the so-called standard quantum limit (SQL). Comparatively, with an optimally entangled probe state, the QFI exhibits a quadratic scaling with $N$, approaching the Heisenberg limit (HL). For the informed environment, the formalism of encoding Hamiltonian $H$ is known and the optimal probe state can be calculated exactly. Yet in many practical cases, as illustrated in Figure \ref{fig:1}a, $H$ is agnostic, and the channel $S_\theta(\rho)$ acts as a ``black-box''. Therefore one has to derive the optimal probe state from information gained from $S_\theta(\rho)$. Such information can be obtained by an oracle $U_{\mathcal{F}}$, which can be a quantum or classical algorithm to estimate the QFI or relevant quantities, as shown in Figure \ref{fig:1}b.

For the optical quantum sensing, we confine the probe state's preparation to a sequence derived from a linear optical toolbox encompassing spontaneous parametric down-conversion (DC), beam splitter (BS), polarized beam splitter (PBS), half-wave plate (HWP), quarter-wave plate (QWP), and mirror reflection (R). Moreover, without loss of generality, we distinguish devices applying on different photon paths or with different parameters. Given the connectivity and noise constraints, the sequence's length is limited. Therefore, to minimize parameter estimation uncertainty, a fundamental requirement is identifying an optimal optical setup for the probe state preparation. For more details about optical experimental devices, refer to supplementary material (SM).

To formalize the search for an optimal optical setup, let us define the toolbox of optical devices as a set $\mathcal{T}$, and the devices correspond to the elements in it. Consequently, an optical setup comprising $L$ devices can be represented as $\mathrm{E}_{enc} \in \mathcal{T}^L$. Hence, the search task converges to the following
\begin{equation}
    \max_{\mathrm{E}_{enc} \in \mathcal{T}^L} \mathcal{F}_Q (\mathrm{E}_{enc}(\rho_{in})),
\end{equation}
which is locating the $\mathrm{E}_{enc}$ within $\mathcal{T}^L$ that maximizes the corresponding QFI $\mathcal{F}_Q(\rho)$. Here we do not pose any constraints on $\mathrm{E}_{enc}$, the space encompassing all possible setups is $|\mathcal{T}|^L$ given a maximum length $L$. Note that $|\mathcal{T}|$ may also grow with the number of qubits $N$. 

Our Deep learning-based Quantum Sensing (DQS) scheme aims to find the optimal optical setups that maximize the QFIs as well as estimate the correct parameter of interest. As shown in Figure \ref{fig:1} c-f, the scheme of DQS comprises four stages: training, ranking, fine-tuning, and response inference. We delve into these stages in the following.

In the training stage, each optical setup is mathematically defined as a directed acyclic graph $G(V,E)$ with $V$ denoting nodes (devices) and $E$ edges (connection). The devices are denoted by an array of node features $\mathbf{X} = (\mathbf{x}_1, \dots, \mathbf{x}_l, \dots, \mathbf{x}_L)^\intercal \in \mathbb{R}^{L \times d}$, and the connection is represented by an adjacency matrix $\mathbf{A} \in \{0,1\}^{L \times L}$, such that $\mathbf{A}_{ij} = 1$ only if the $j$-th device is connected to the $i$-th device. To extract graph-level feature, we employ a GNN that maps the devices $\mathbf{X}$ and the connection $\mathbf{A}$ to a latent vector $\mathbf{z} \in \mathbb{R}^s$. Subsequently, a Multilayer Perceptron (MLP) operates on $\mathbf{z}$ to predict the QFI. The training dataset, denoted by $\{X_i, A_i, \mathcal{F}(X_i,A_i) \}_{i=1}^D$ where $\mathcal{F}(X_i,A_i)$ is the label, is generated via the oracle $U_\mathcal{F}$ as illustrated in Figure \ref{fig:1}b. The entire deep learning model, $\mathcal{G}$, is trained in a supervised manner with the objective of minimizing the prediction error $\mathcal{L}$. Mathematically, this objective is expressed as:
\begin{equation}
    \min_{\mathbf{W}\in \mathcal{W}} \mathcal{L}(\mathbf{W})=\frac{1}{D}\sum_{i=1}^D \left( \mathcal{G}(X_i, A_i)- \mathcal{F}(X_i)\right)^2,
\end{equation}
where $\mathbf{W}$ is the weight of neural network $\mathcal{G}$. For more details of the implementation of training, refer to SM.

Post-training, the model $\mathcal{G}$ remains fixed and is used to predict the QFIs of a set of offline explored, unlabeled setups. As illustrated in Figure \ref{fig:1}d, these predictions are subsequently organized and ranked. Based on this ranking, setups exhibiting the highest QFI are selected as prime candidates $(\mathbf{X}^*, \mathbf{A}^*)$. For instance, we might extract samples associated with the top-K QFIs for subsequent stages.

The candidates identified during ranking undergo further validation and evaluation. Moreover, these setups are optimized, as shown in Figure \ref{fig:1}e, by eliminating redundant devices or adjusting their positions where feasible. The most effective setup is then selected to prepare the final probe state.

When deploying the final probe state to interact with the unknown Hamiltonian, the response function $R(\theta)$ has no explicit definition. To this end, as shown in Figure \ref{fig:1}f, we employ a trigonometric polynomial of degree $n$ to infer the response function, expressed as 
\begin{equation}
    R(\theta)=\sum_{s=1}^N [a_s \cos(s\theta)+b_s \sin(s\theta)]+c.
\end{equation}
Using the $2N+1$ predefined parameters $\{\theta_k\}_{k=1}^{2N+1}$, the coefficients can be approximated via trigonometric interpolation \cite{alderete2022inference}. Thereafter, for a new $\theta$ and its associated measurement output $\bar{R}$, the inferred value of $\theta$ is determined as $\tilde{\theta} = \argmin_{\theta}|\tilde{R}(\theta)- \bar{R}|$ where $\tilde{R}(\theta)$ is the inferred response. For an in-depth explanation, refer to the SM.

\begin{figure*}[th]
\centering
\includegraphics[width=0.98\textwidth]{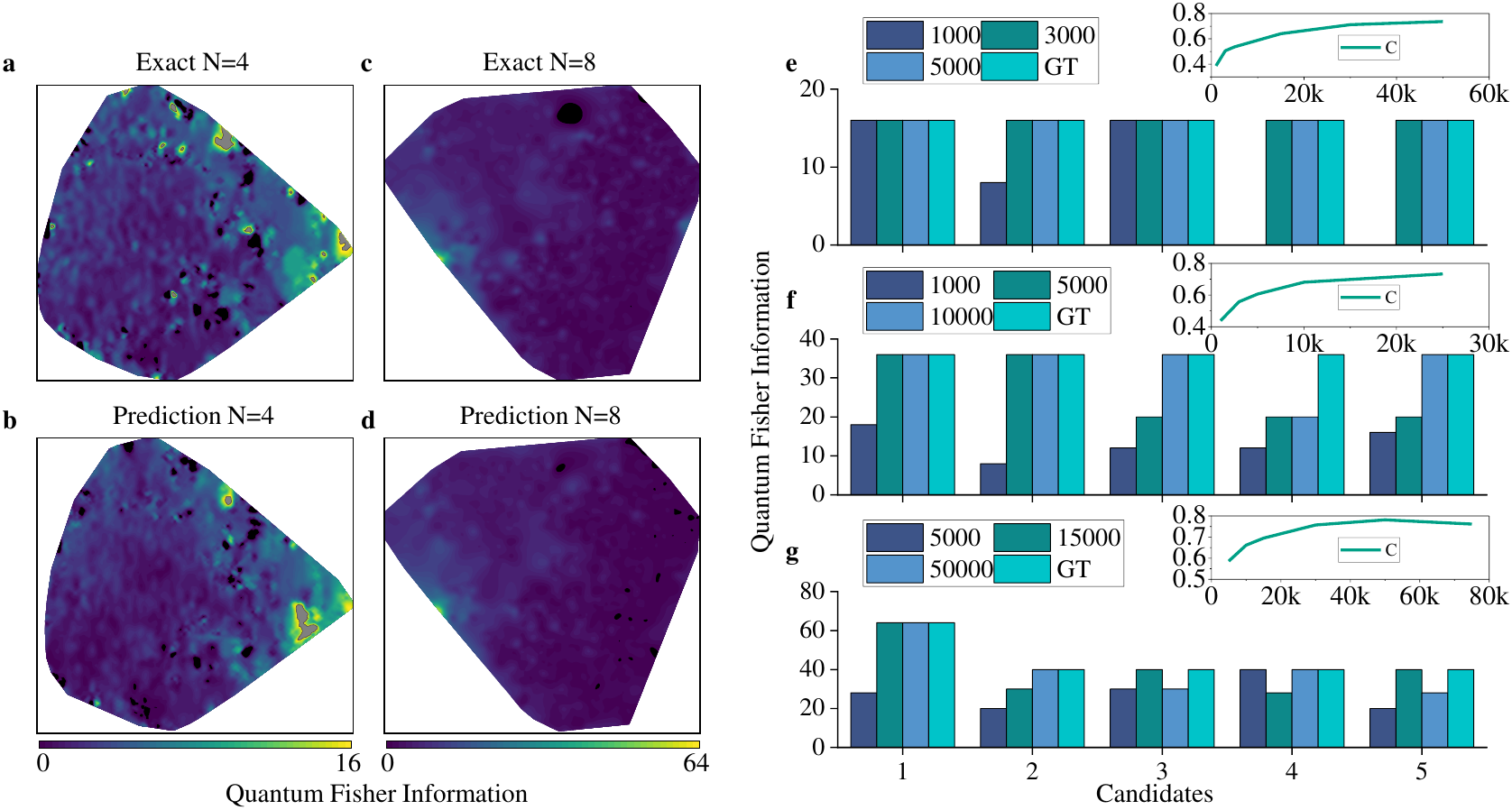}
\caption{\textbf{Performance of DQS on different datasets.} \textbf{a-d. Visualization of the Latent Space of optical setups.} \textbf{a} and \textbf{c} depict the latent spaces for 4-photon and 8-photon setups, respectively. The color bar indicates the Quantum Fisher Information (QFI) levels. As evident in these figures, the Graph Neural Network (GNN) maps each setup to a corresponding latent vector based on its QFI, clustering higher QFI values toward the corners of the latent space. \textbf{b} and \textbf{d} show GNN's QFI predictions in the respective latent spaces. \textbf{e-g. Performance Analysis in Relation to the Size of the Training Data.} The primary bar chart illustrates the top 5 setups selected from 10,000 random samples, based on models trained with varying data sizes, alongside the ground truth (GT) top 5. Subfigures \textbf{e} through \textbf{g} correspond to tasks involving 4, 6, and 8 photons. The inset graph elucidates the relationship between the Spearman correlation coefficients (C) and the sizes of the training data sets.}
\label{fig:2}
\end{figure*}
\section{Numerical Simulations}
\textbf{Performance of GNN predictor}.  To showcase the capabilities of our DQS algorithm, we begin by analyzing the behavior of the GNN in the latent space. We consider both 4-photon and 8-photon tasks, where $H=\sum_{i=1}^N Z_i$ and the maximum setup length is 15. The GNN feature extractor transforms each setup into a 256-dimensional vector. Due to the complexities in visualizing high-dimensional data, we utilize the t-SNE method for dimensionality reduction, compressing the data into two dimensions. In Figure \ref{fig:2}a and c, data points of 10000 test samples in the latent space are color-coded according to their respective QFIs. The GNN successfully processes setups of various lengths and layouts, originally in disparate data formats, into uniform dimensional vectors, thereby facilitating subsequent analyses. Furthermore, the GNN clusters setups with higher QFIs within its latent space; for example, the right corner in the 4-photon latent space and the left corner in the 8-photon space. From an information-processing perspective, the GNN retains relevant QFI details in its latent space while filtering out irrelevant data. Notably, the area with higher QFIs in the 4-photon latent space is considerably larger than in the 8-photon space, reflecting the increasing difficulty of locating optimal probes as the number of qubits grows.

Next, we turn our attention to the MLP predictor. Figure \ref{fig:2}b and d present the prediction results in the same latent spaces depicted in a and c. The predicted QFIs closely align with the ground truth distribution. Importantly, even though the predicted QFI exhibits a small deviation, this does not impact the ranking phase, as the order of QFIs remains consistent within the highlighted regions of high-QFI setups.

Finally, we evaluate the GNN model's performance in relation to the size of the training data set. The aim is to identify the top-5 candidate setups, according to the QFI, from a pool of 10,000 random test samples in each task. Figure \ref{fig:2}e-g, corresponding to 4-photon, 6-photon, and 8-photon tasks, displays the QFIs of the top-5 setups as suggested by GNN models trained on varying data sizes, in comparison to the exact top-5 setups in 10,000 test samples. The corresponding QFIs of the optimal probes in these tasks are 16, 36, and 64 respectively. The results underscore the importance of data size. For instance, in the 4-photon task, the model trained on only 1,000 data points includes two setups with a QFI of zero among its top 5, which is clearly not optimal. Similarly, in the 6-photon task, the model trained on 1,000 data points fails to identify the optimal setup, whereas the model trained on 5,000 data points finds two, and the one trained on 10,000 finds four. As data size increases, so does the number of optimal setups among the top 5 candidates. We employ the Spearman correlation coefficient to illustrate the improvement in ranking quality as a function of increasing data size, as shown in the inset of Figure \ref{fig:2}e-g. The required volume of data increases in proportion to the number of qubits, as the size of the search space expands. Accordingly, the GNN model requires more extensive datasets to ensure robust performance on unseen samples.

\begin{figure*}[ht]
\centering
\includegraphics[width=0.98\textwidth]{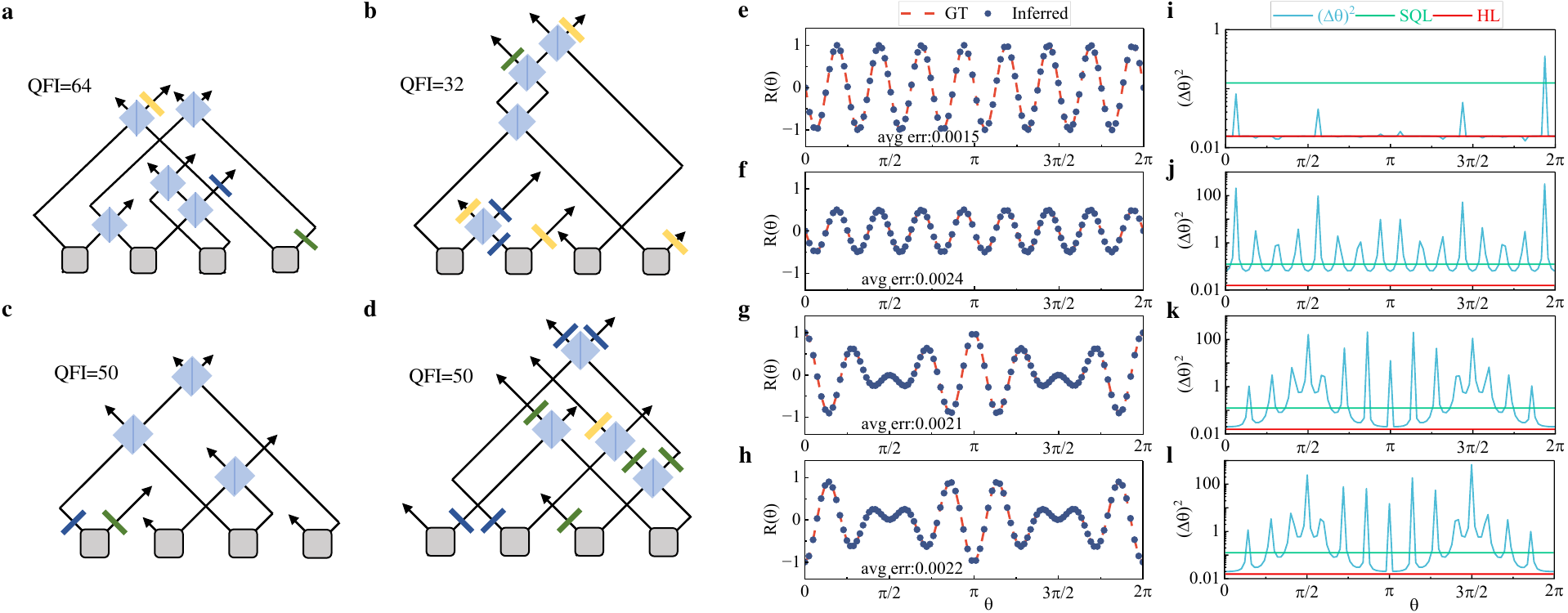}
\caption{
\textbf{Simulated results of an eight-photon quantum sensing task.}
\textbf{a-d. Experimental Layout.}
\textbf{a-d} presents the layout of the top-3 probe state identified during the testing phase by our method, while \textbf{d} shows the layout found within the training samples. The optical devices are represented as follows: DC (gray square), PBS (blue square), HWP (yellow rectangle), QWP (green rectangle), and R (deep blue rectangle).
\textbf{e-h. Inferred Response Functions.}
Subfigures \textbf{e-h} display the inferred response functions corresponding to the layout shown in \textbf{a}-\textbf{d}. The dashed red line represents the exact response function (GT), while the points indicate the inferred response.
\textbf{i-l. Estimated Sensitivity.}
Subfigures \textbf{i-l} exhibit the estimated sensitivities related to the layout depicted in \textbf{a}-\textbf{d}. The blue curve indicates the sensitivity level, while the green and red lines mark the SQL and HL, respectively.
}
\label{fig:3}
\end{figure*}

\textbf{Optimal Eight-Photon Quantum Sensing}. We now turn our attention to an 8-photon quantum sensing task to investigate the overall performance for parameter estimation. In this context, we consider $H=\sum_{i=1}^8 Z_i$ and $O=\otimes_{i=1}^8 X_i$, which are unknown to our model. The GNN model under evaluation has been trained on 30,000 samples and is subsequently used to explore 50,000 new samples. We identify the top-3 candidates, as illustrated in Figure \ref{fig:3}a-c, providing the final state $|\psi\rangle$ and QFI as follows: (1) $|\psi\rangle=((1 - i)|0\rangle^{\otimes8}+ (1 + i)|1\rangle^{\otimes8})/2$ and QFI being 64; (2)  $|\psi\rangle=\frac{-1+i}{\sqrt{2}}(|0\rangle^{\otimes4}+i|1\rangle^{\otimes4})(|0\rangle^{\otimes4}+|1\rangle^{\otimes4})$ and QFI being 32; (3)  $|\psi\rangle=(i|0\rangle_b+|1\rangle_b)(|0\rangle^{\otimes7}+|1\rangle^{\otimes7})$ and QFI being 50. The number of devices is 8, 11, and 5 respectively. For comparative analysis, we also provide the best optical setup identified within the training data, as depicted in Figure \ref{fig:3}d, which serves as a result of the exhaustive search: $|\psi\rangle=(\frac{1-i}{\sqrt{2}}|0\rangle_g+\frac{i-1}{\sqrt{2}}|1\rangle_g)(|0\rangle^{\otimes7}+|1\rangle^{\otimes7})$ and QFI being 50. The number of devices is 13. Here, the subscripts indicate the device parameters and applied photon paths (from $a$ to $h$). For the detailed device sequences of optical setups, refer to the SM.

The first candidate is indeed an 8-qubit Greenberg-Horne-Zeilinger (GHZ) state, while others are tensor products of local GHZ states. The QFI of 64 indicates that our model finds the optimal optical setup in probe preparation. Notably, the probe state prepared by Figure \ref{fig:3}d exhibits a QFI of 50, meaning that none of the training samples possess a QFI of 64 as \ref{fig:3}a. Despite the lack of exposure to the optimal setup during training, our model still manages to identify it among the unlabeled samples in the testing phase. This result highlights that the GNN model is in fact learning, rather than memorizing by rote, key structural patterns that contribute to QFI maximization, such as optimal utilization of beam splitters or wave plates, in an implicit way.

Subsequently, using these four prepared probes, the measurement output $\bar{R}(\theta)$ is collected, averaging over 10,000 shots results for each $\theta \in \{\theta_k\}_{k=1}^{2n+1}$ where $\theta_k = \frac{2\pi(k-1)}{2n+1}$. We employ the discussed trigonometric interpolation technique to approximate the response functions, as depicted in Figure \ref{fig:3}e-h. The optimal probe state prepared using our method yields response functions with minimal errors (Figure \ref{fig:3}e, average error 0.0015), compared to the exhaustive search result (Figure \ref{fig:3}h, average error 0.0022).  Interestingly, we observe error amplification at points with minimal derivatives, reflected by the concentration of points in small derivative intervals. This is consistent with the expectation that distinguishing between distinct $\theta$ values becomes challenging in regions of flat response.

Finally, we estimate the sensitivity of the whole quantum sensing scheme, as formulated in
\begin{equation}
    (\Delta \theta)^2 = \frac{1-(\sum_{s=1}^n [a_s \cos(s\theta)+b_s \sin(s\theta)]+c)^2}{|\sum_{s=1}^n s[-a_s \sin(s\theta)+b_s \cos(s\theta)]|^2}.
\end{equation}
We compare this estimated sensitivity against both the SQL (0.125) and the HL (0.016), as illustrated in Figure \ref{fig:3}i-l. Our proposed method closely approximates the HL across the majority of the interval, shown in Figure \ref{fig:3}i, with minor deviations attributed to finite measurement shots. However, the best-performing setup from the training set, shown in Figure \ref{fig:3}l, only approximates the HL within small intervals and frequently falls short of even the SQL. These results demonstrate that, with the same estimation method, the sensitivity of the quantum sensing scheme can still vary and even be inferior to classical sensors due to the suboptimal probe preparation. Regardless of the agnostic environment, our proposed DQS scheme manages to prepare the optimal probe and reach the HL in terms of sensitivity.

\section{Discussion and Outlook}
Our investigation illuminates the potential of deep learning in optical quantum sensing, especially in scenarios where traditional scheme design falls short due to unknown target systems. The DQS scheme we have developed overcomes the challenge by offering an algorithmic scheme that not only learns to identify the optimal probes but also to estimate the target parameter to the Heisenberg limit (HL). The integrated GNN adeptly distills informative representations from the configurations of optical experiments, thereby amplifying the algorithm's efficacy in identifying optimal optical setups. Numerical experiments demonstrate the efficiency of our DQS method and its ability to evolve with increasing data. In a simulated eight-photon quantum sensing task, DQS achieves lower parameter estimation error compared to trivial quantum sensing scheme, and approaches HL in precision analysis. In a growing field where quantum technologies are becoming both more advanced and more complex, a scalable, data-driven approach for enhancing quantum sensing stands as a crucial advancement. As such, our work acts as a pivotal link between deep learning and quantum sensing, highlighting a pathway for accelerated advancements in practical quantum technologies.

Despite the strides made with our DQS in enhancing the discovery of optimal quantum sensing schemes, certain limitations persist. Firstly, our research scope did not encompass the design of the measurement operator. We exclusively utilized a basic Pauli operator, neglecting its pivotal role in achieving HL \cite{giovannetti2011advances}. When delving into unknown environments, an intriguing avenue for future exploration is the development of adaptive measurement operators \cite{bonato2016optimized,marciniak2022optimal}. This would align the measurement with its probe and environment, potentially integrating into the DQS framework, thereby synchronizing the design for both probes and measurements. Secondly, accessing a quantum oracle to estimate the QFI remains a complex endeavor \cite{liu2020quantum}. This is closely tied to accumulating training examples. Anticipated future research could explore efficient QFI data collection methodologies or even consider training deep learning models with partial measurement results as label \cite{demkowicz2020multi,yu2021experimental,rath2021quantum,gacon2021simultaneous}. Lastly, our DQS strategy currently simply ranks random samples and picks the best candidates after the training phase. A potential enhancement could merge our model with extant optimization algorithms, such as evolutionary algorithms \cite{krenn2016automated, knott2016search, o2019hybrid, nichols2019designing}, further refining candidate searches. These future investigations will help build a more robust deep learning-based quantum sensing scheme.

Beyond quantum sensing and optics, our research serves as a catalyst for novel explorations in harnessing deep learning techniques to learn and predict quantum systems \cite{torlai2018neural,zhu2022flexible,koutny2023deep}. Two salient questions emerge: Firstly, how can quantum system-generated data be optimally represented for more effective learning? In DQS, we employ a graph format for depicting an optical setup, capturing both device data and layout dynamics. In contrast, outputs like shadow tomography \cite{aaronson2018shadow,huang2020predicting} might better fit array representations and be used for certification task \cite{du2023shadownet}. Secondly, how can deep learning models be custom-crafted for specific quantum systems? GNNs may be better suited for graph-structured data, like quantum circuits, over sequential data typified by measurement outputs. Moreover, the distinct characteristics of quantum-generated data necessitate strategic model training, warranting additional exploration.

\emph{Acknowledgement.} We thank Xuemei Gu for helpful discussions on optical quantum experiments. This work was supported in part by NSFC No. 62222117. X.-F. Y. acknowledges support from the China Postdoctoral Science Foundation (Grant No. 2023M733418 )

%

\appendix

\renewcommand{\thefigure}{M\arabic{figure}}	
\setcounter{figure}{0}

\section{Optical quantum computing}
\subsection{Quantum State Representation}
We discuss qubits that are encoded through photon polarization. Specifically, the state $|0\rangle$ corresponds to vertically polarized light, while $|1\rangle$ relates to horizontally polarized light. In this representation, an n-qubit quantum state (or equivalently, an n-photon state) can be written as:
\begin{equation}
    |\psi\rangle = \sum_{i\in \{0,1\}^n} \alpha_i |i\rangle.
\end{equation}
Compared to other quantum platforms, optical quantum computing has some unique phenomena, such as the HOM effect. This effect causes several photons to occupy the same path, making them indistinguishable, and effectively reducing them to a single-qubit state. We use post-processing to ensure that an n-qubit state remains a superposition of n-qubit bases. For instance, when two photons are equally likely to be in one path or to split between two paths, the state is represented as  $|\psi\rangle = |0_a0_b\rangle + |0^2_{a}\rangle$ (where the superscript 2 indicates two indistinguishable photons in path a). Post-processing then simplifies this state to $|\psi\rangle = |0_a0_b\rangle$.

\subsection{Initial state and linear optical devices}

\begin{table*}[h]
\centering
\begin{tabular}{c|c|c|c|c}
\hline
\textbf{Device} & \textbf{Token} & \textbf{Visual} & \textbf{Operation} & \textbf{Operator} \\
\hline
Spontaneous Parametric Down Conversion & $\text{DC}(|\psi\rangle, p, p^\prime)$ & \adjustbox{valign=m}{\includegraphics[width=0.7cm, height=0.7cm]{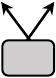}} & $|\psi\rangle \otimes \sum_{l}|l\rangle_p|l\rangle_{p^\prime}$ & $\text{DC}_{p,p^\prime}$ \\ 
\hline
Beam Splitter & $\text{BS}(|\psi\rangle, p, p^\prime)$ & \adjustbox{valign=m}{\includegraphics[width=0.7cm, height=0.7cm]{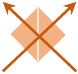}} & \begin{tabular}[c]{@{}c@{}}$|l\rangle_p \rightarrow \frac{|l\rangle_p + i|l\rangle_{p^\prime}}{\sqrt{2}} $\\ $|l\rangle_{p^\prime} \rightarrow \frac{|l\rangle_{p^\prime} + i|l\rangle_{p}}{\sqrt{2}} $\end{tabular}   & $\text{BS}_{p,p^\prime}$ \\ 
\hline
Polarized Beam Splitter & $\text{PBS}(|\psi\rangle, p, p^\prime)$ & \adjustbox{valign=m}{\includegraphics[width=0.7cm, height=0.7cm]{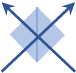}} & \begin{tabular}[c]{@{}c@{}}$|l\rangle_p \rightarrow |0\rangle_p \ \text{when}\ l=0 $\\ $|l\rangle_{p} \rightarrow |1\rangle_{p^\prime} \ \text{when}\ l=1 $\end{tabular}   & $\text{PBS}_{p,p^\prime}$ \\ 
\hline
Half Wave Plate & $\text{HWP}(|\psi\rangle, p, \theta)$ & \adjustbox{valign=m}{\includegraphics[width=0.7cm, height=0.7cm]{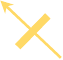}} & \begin{tabular}[c]{@{}c@{}}$|l\rangle_p \rightarrow O_{\text{HWP}}|l\rangle_p $\\ $ O_{\text{HWP}}(\theta)=\begin{bmatrix}
    \cos 2\theta & \sin 2\theta \\
\sin 2\theta & -\cos 2\theta \\
\end{bmatrix} $\end{tabular}   & $\text{HWP}_{p,\theta}$ \\ 
\hline
Quater Wave Plate & $\text{QWP}(|\psi\rangle, p, \theta)$ & \adjustbox{valign=m}{\includegraphics[width=0.7cm, height=0.7cm]{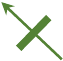}} & \begin{tabular}[c]{@{}c@{}}$|l\rangle_p \rightarrow  O_{\text{QWP}}|l\rangle_p $\\ $ O_{\text{QWP}}(\theta)=\frac{1}{\sqrt{2}}\begin{bmatrix}
    1-i\cos 2\theta & -i\sin 2\theta \\
-i\sin 2\theta & 1+i\cos 2\theta \\
\end{bmatrix} $\end{tabular}   & $\text{QWP}_{p,\theta}$ \\ 
\hline
Mirror Reflection & $\text{R}(|\psi\rangle, p)$ & \adjustbox{valign=m}{\includegraphics[width=0.7cm, height=0.7cm]{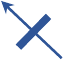}} & $|l\rangle_p\rightarrow i|l\rangle_p$ & $\text{R}_{p}$ \\ 
\hline
\end{tabular}
\caption{Toolbox of devices.}
\label{table:1}
\end{table*}

Initial quantum states are produced via a process known as spontaneous parametric down conversion (SPDC, here we use DC for short). For instance, applying DC on paths `a' and `b' can produce states such as $|0_a0_b\rangle$, $|1_a1_b\rangle$, or a superposition $|0_a0_b\rangle+|1_a1_b\rangle$. To generate a multi-qubit initial state, one can use multiple DC processes. For example, to get a 4-qubit initial state, one might produce $|0_a0_b\rangle\otimes|1_c1_d\rangle$. Once the initial state is produced using DC, various devices (referenced in Table \ref{table:1}) act on it to prepare the desired quantum state.

\begin{figure*}[ht]
\centering
\includegraphics[width=0.98\textwidth]{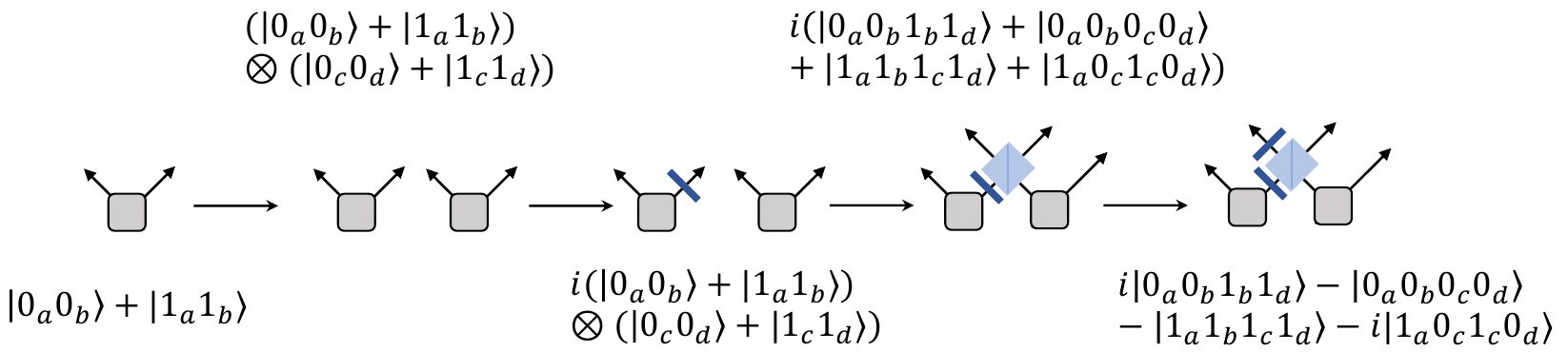}
\caption{Example of preparation of quantum state.}
\label{fig:optcs}
\end{figure*}

In Figure \ref{fig:optcs}, we show how a quantum state is generated with an optical setup. Starting with two DCs to prepare the initial state, the sequence ``$\mathrm{R}_b\rightarrow \mathrm{PBS}_{b,c}\rightarrow\mathrm{R}_c$'' is applied on it. The state is $i|0_a 0_b 1_b 1_d \rangle-|0_a 0_b 0_c 0_d \rangle-|1_a 1_b 1_c 1_d \rangle-i|1_a 0_c 1_c 0_d\rangle$. After post-processing to filter out the invalid basis, the state is $-|0_a 0_b 0_c 0_d \rangle-|1_a 1_b 1_c 1_d \rangle$, which is a GHZ state.

\section{Optical quantum sensing}

\subsection{Framework}
We consider single-parameter estimation quantum sensing. In this approach, we utilize an $n$-qubit probe state, denoted as $\rho$, to estimate an unknown parameter $\theta$. This parameter is encoded using a unitary channel in the form
\begin{equation}
    S_\theta(\rho) = e^{-\frac{1}{2}i\theta H}\rho e^{\frac{1}{2}i\theta H},
\end{equation} 
where $H$ represents the encoding Hamiltonian such that $H = \sum_j h_j$ with $h_j^2=I$ and $[h_j,h_{j^\prime}]=0,\, \forall j,j^\prime$. After measuring the expectation value of an observable $O$, which conforms to the constraint $\|O\|_{\infty}\leq 1$, the response function is defined as 
\begin{equation}
    R(\theta) = \mathrm{Tr}[S_\theta (\rho) O].
\end{equation}
To illustrate, given $H=\sum_{i=1}^N Z_i$ and $O=\otimes_{i=1}^N X_i$, as well as the probe state $\rho = |\psi\rangle\langle\psi|$ with $|\psi\rangle = (|0\rangle^{\otimes N} + e^{i\gamma}|1\rangle^{\otimes N})/\sqrt{2}$, the response function is $R(\theta)^2=\cos N\theta$.  

\subsection{Fisher Information and Cramer-Rao bound}

In an experiment, the measurement outcome $x$ is influenced by the parameter $\theta$, and the conditional probability distribution of $x$ for a given $\theta$ is denoted as $p(x|\theta)$. The Fisher information \cite{fisher1922mathematical} for the parameter $\theta$ is defined as:
\begin{equation}
    F(\theta) = \mathbb{E}\left[\left(\frac{\partial \log p(x|\theta) }{\partial\theta}\right)^2\right],
\end{equation}
which quantifies the sensitivity of $x$ to changes in $\theta$. For an unbiased estimator $\tilde{\theta}$ of $\theta$ derived from measurement results $\{x_1,\dots,x_M\}$ over $M$ experiments, the Cramer-Rao bound (CRB) \cite{rao1992information} holds:
\begin{equation}
    (\Delta\theta)^2\geq \frac{1}{MF(\theta)}.
\end{equation}

\subsection{Quantum Fisher Information and Quantum Cramer-Rao bound}

In a quantum context, the sensitivity of $x$ is inherently associated with the measurement process. For a parameterized quantum state $\rho_\theta$, the Fisher information can be maximized over all possible POVMs $E_x$, leading to the quantum Fisher information (QFI) \cite{braunstein1994statistical}:
\begin{equation}
    F_Q(\rho_\theta) = \max_{\{E_x\}}F(\theta).
\end{equation}
Subsequently, the Quantum Cramer-Rao bound (QCRB) is
\begin{equation}
    (\Delta\theta)^2\geq \frac{1}{MF(\theta)} \geq \frac{1}{MF_Q(\rho_\theta)}.
\end{equation}
Given a fixed probe state, this bound indicates the ultimate precision regardless of measurement.

By introducing the symmetric logarithmic derivative operator $L_\theta$ as
\begin{equation}
    \frac{\partial\rho_\theta}{\partial \theta} = \frac{1}{2}(L_\theta\rho_\theta+\rho_\theta L_\theta),
\end{equation}
it can be demonstrated that 
\begin{equation}
    F_Q(\rho_\theta) = \mathrm{Tr}(\rho_\theta L_\theta^2).
\end{equation}

If the evolution of interaction is unitary, i.e. $\rho_\theta = e^{i\theta H}\rho e^{-i\theta H}$, $F_Q$ does not depend on $\theta$. Given the Hamiltonian $H$ and the probe state $\rho = \sum_nb_n|\Phi_\rangle\langle\Phi_n|$, the explicit form of $F_Q$ is:
\begin{equation}
    F_Q = 2\sum_{i\neq j}(\frac{(b_i-b_j)^2}{b_i+b_j}|\langle\Phi_i|H|\Phi_j\rangle|).
\end{equation}
For a pure state, namely $\rho = |\Psi\rangle\langle\Psi|$, a simpler expression is
\begin{equation}
    F_Q = 4(\Delta H)^2,
\end{equation}
where $(\Delta H)^2 = \langle(H-\langle H \rangle)^2\rangle$. 

For probe states with classical correlations, such as a tensor product of single-qubit states, the QFI scales linearly with the qubit number $N$. However, for optimal entangled probe states, the QFI scales quadratically with $N$, which is related to Heisenberg limit and gives the quantum advantage in sensing precision.

To sum up, the estimator of the parameter needs to be optimal to saturate the quantum Cramer-Rao bound (QCRB). And to make QCRB reach the ultimate Heisenberg limit, the probe state also needs to be optimal. With both requirements satisfied, the quantum sensing is optimal.

\section{Implementation of Deep learning-based Quantum Sensing  scheme}

\subsection{Graph encoding of optical setup}

\begin{figure*}[ht]
\centering
\includegraphics[width=0.98\textwidth]{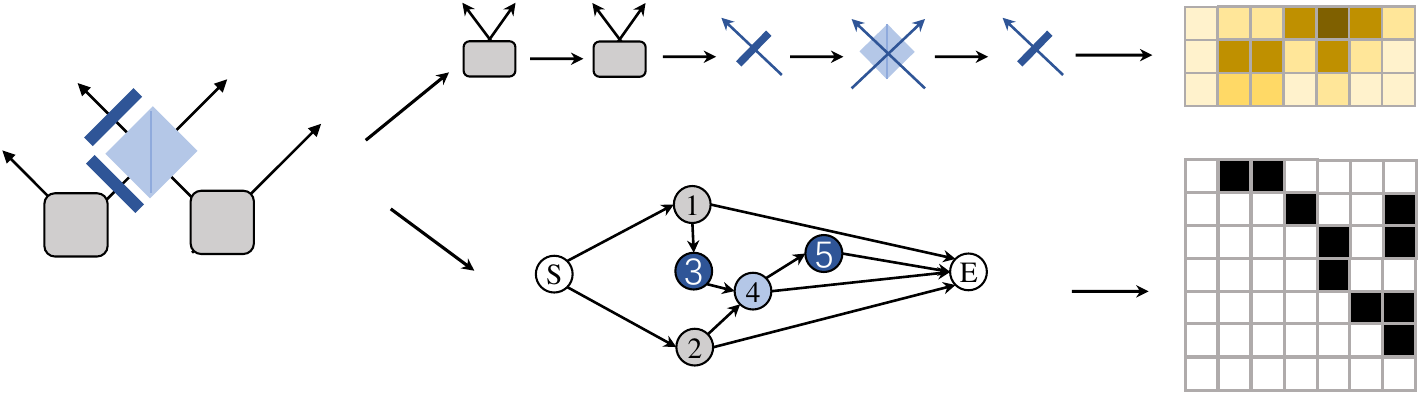}
\caption{Example of graph encoding of optical setup.}
\label{fig:sm1}
\end{figure*}

We delve into the graph encoding of optical setups, as illustrated in Figure \ref{fig:sm1}. A setup can be characterized by the sequence of devices and their topological connections.

The sequence depicts the device order, which we represent as a matrix $\mathbf{X}$. This matrix is defined as  $\mathbf{X} = (\mathbf{x}_1, \dots, \mathbf{x}_l, \dots, \mathbf{x}_L)^\intercal \in \mathbb{R}^{L \times d}$, as shown in Figure \ref{fig:sm1}c. Each $\mathbf{x}_i$ within the matrix is a vector that represents the corresponding i-th device in the sequence and is known as the feature vector. This vector is binary and consists of two components. The first part is a one-hot encoded vector (of length $d_1$) that identifies the device type. In this encoding, only one element is 1, while the others are 0. It is noteworthy that devices with varying parameters are distinguished.  For instance, a QWP with $\theta=\frac{\pi}{4}$ and another with $\theta=\frac{\pi}{2}$ are considered different devices. In the main text, we consider a quantization of $\frac{\pi}{4}$ so there are four types of HWP and four types of QWP. Besides, we also treat DC with initial state $|00\rangle$, $|11\rangle$, and $|00\rangle+|11\rangle$ as three devices. Without any loss of generality, we also incorporate two unique devices: the start and end indicators. The second component is positional encoding. Here, each bit in the vector corresponds to a path. If a device interacts with path `a', then the associated bit in the vector (of length $d_2$) is set to 1; otherwise, it is set to 0.

The topological connection is defined using a directed acyclic graph (DAG) $G(V,E)$ with $V$ denoting the node set and $E$ the edge set. Every node corresponds to a device. An edge $e_{ij}$ exists only if the j-th device functions on the same path subsequent to the i-th device. Every DC node is connected after the start node, and every node ends up with either another device node or the end node. Here, we employ the adjacency matrix $\mathbf{A}\in\{0,1\}^{L\times L}$ to represent $G$. In this matrix, $\mathbf{A}_{ij}=1$ if and only if $e_{ij}\in E$.

\subsection{Graph neural network}

The Graph neural network used in DQS scheme is structured into four components: embedding, graph transformer (message passing function), max-pooling (READOUT function), and predictor. In the following, we detail the implementation of these four components. For clarification,  all the $\mathbf{W}$ and $\mathbf{b}$ variables denote the adjustable weights of the neural network. Additionally, ``Act'' represents the activation function, and ``BN'' denotes the batch normalization function. The network is implemented using PyG library \cite{fey2019fast}.

\begin{figure}[h]
\centering
\includegraphics[width=0.48\textwidth]{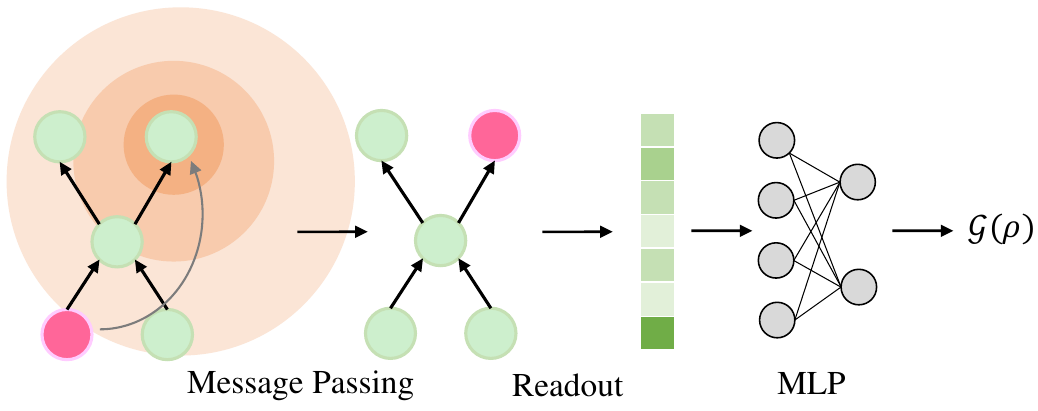}
\caption{\textbf{Graph Neural Network.} The information flows between node neighbors through the Message Passing stage. After the Readout, the latent feature is sent to MLP to produce the final prediction $\mathcal{G(\rho)}$. }
\label{fig:gnn}
\end{figure}

\textbf{Embedding Phase}. An optical setup is modeled as a graph $G(V,E)$, where every device within it corresponds to a node and is encapsulated as a vector $\mathbf{x}_v \in \{0,1\}^{d_1 + d_2}$, in which the initial $d_1 = |\mathcal{T}|$ bits serve as a one-hot encoding for the device type, while the remaining $d_2 = N$ bits function as the positional encoding of the device. The directed edges correspond to the paths of the photon. Typically, GNNs follow an iterative neighborhood aggregation scheme to capture the information within the nodes' neighborhood. The embedding phase uses a single-layer MLP to transform the node feature vector, aligning it with the latent space. This transformation is given by:
\begin{equation}
\mathbf{h}_i^{(0)} = \mathrm{MLP}_0(\mathbf{x}_i) = \mathrm{Act}(\mathrm{BN}(\mathbf{W}_0\mathbf{x}_i+\mathbf{b})),
\end{equation}
where the dimensions of $\mathbf{x}_i$ and $\mathbf{h}_i^{(0)}$ are $d$ and $s$. 

\textbf{Graph Transformer}.  Following the embedding phase, we employ alternating layers of graph transformers (GT) \cite{shi2020masked} and MLPs to serve as the message passing mechanism. The $l$-th is defined by:
\begin{equation}
    \textbf{h}_i^{(l)} = \mathrm{MLP}_l(\mathrm{GT}_l(\textbf{h}_i^{(l-1)})).
\end{equation}
Here, a C-head graph transformer is represented as
\begin{equation}
    \hat{\textbf{h}}_i^{(l)} = \mathrm{GT}(\textbf{h}_i^{(l-1)})= \mathrm{Concat}(\hat{\textbf{h}}_i^{(l,1)},\dots,\hat{\textbf{h}}_i^{(l,C)}).
\end{equation}
The output of the k-th head can be defined by
\begin{equation}
    \hat{\textbf{h}}_i^{(l,k)} = \beta_i^{(l,k)}\mathbf{W}_1^{(l,k)} \textbf{h}_i^{(l-1)} + (1-\beta_i^{(l,k)})\mathbf{m}_i^{(l-1,k)},
\end{equation}
using aggregation
\begin{equation}
    \mathbf{m}^{(l,k)}_i = \sum_{j\in \mathcal{N}(i)} \alpha_{i,j,k}^{(l)} \mathbf{W}_2^{(l,k)}\mathbf{h}_j^{(l-1)},
\end{equation}
and attention coefficients
\begin{equation}
    \alpha_{i,j,k}^{(l)} = \mathrm{softmax}(\frac{(\mathbf{W}_3^{(l,k)} \textbf{h}_i^{(l-1)})^{\intercal}(\mathbf{W}_4^{(l,k)} \textbf{h}_j^{(l-1)})}{\sqrt{s}}),
\end{equation}
alongside the skip information factor
\begin{equation}
    \beta_i^{(l,k)} =   \mathrm{sigmoid}({\mathbf{w}_5^{(l,k)}}^\intercal [\mathbf{W}_1^{(l,k)} \textbf{h}_i^{(l-1)}, \mathbf{m}^{(l,k)}_i,  \mathbf{W}_1^{(l,k)} \textbf{h}_i^{(l-1)}- \mathbf{m}^{(l,k)}_i] ).
\end{equation}
The MLP is defined by
\begin{equation}
    \textbf{h}_i^{(l)} = \mathrm{MLP}(\hat{\textbf{h}}_i^{(l)} ) = \mathbf{W}_7^{(l)}(\mathrm{Act}(\mathrm{BN}(\mathbf{W}_6^{(l)}\hat{\textbf{h}}_i^{(l)}+\mathbf{b}))) + \mathbf{b}.
\end{equation}

\textbf{Max-pooling}. This phase merges all node vectors $(\textbf{h}_1^{(N)},\dots,\textbf{h}_L^{(N)})$ into an one-dimensional vector $\textbf{h}\in \mathbb{R}^{s}$ by
\begin{equation}
    \textbf{h} = \mathrm{Max}(\textbf{h}_1^{(N)},\dots,\textbf{h}_L^{(N)}),
\end{equation}
where $\mathrm{Max}$ applies element wise.

\textbf{Predictor}. Post-READOUT, a single-layer MLP acts as the neural predictor for QFI
\begin{equation}
    \mathcal{G} = \mathbf{w}^\intercal \mathbf{h} + b.
\end{equation}
Here $\mathcal{G}$ is the prediction of QFI.

\subsection{Training and data}
Our model's architecture is based on five layers of graph convolution, each with a 4-head transformer, with a latent dimension set to 256. The chosen activation function for the model is the GELU function \cite{hendrycks2016gaussian} and the Readout function used in the model is the max pooling function. 

The core objective of supervised learning is to train the neural network with labeled datasets in order to minimize prediction error. In this work, the loss function is given by:
\begin{equation}
    \mathcal{L}_{\mathrm{MSE}}= \mathbf{E} [ \mathcal{G}(\mathbf{X},\mathbf{A})- \mathcal{F}(\mathbf{X},\mathbf{A})]^2,
\end{equation}
where $\mathcal{F}(\mathbf{X})$ is the exact QFI.

Regarding the training process, we utilized the gradient descent optimization technique with the Adam optimizer \cite{kingma2014adam}. We set the learning rate and weight decay at $10^{-4}$ and $10^{-5}$ respectively. Our training consisted of 200 epochs with a batch size of 64. For the implementation of the model and the training process, we employed the PyTorch library \cite{paszke2019pytorch}.

The dataset for optical setups is created using a modification of the Melvin algorithm, a symbolic algebra-based tool  \footnote{https://github.com/XuemeiGu/MelvinPython}.  We want to emphasize that future researchers aiming to replicate or build upon our work are not limited to this particular method. They can use any approach to produce datasets, provided that the resulting data aligns with the graph encoding structure we described.

\subsection{Ranking and Fine-tuning}

In the ranking phase, we randomly sample a substantial amount of optical setups, denoted as $\{X_i,A_i\}_{i=1}^{D^\prime}$. We use a trained GNN model to predict the QFI, represented as $\mathcal{G}(X_i,A_i)$. The results are ranked based on the predicted values, which is
\begin{equation}
    i^* = \argmax_{i \in [D^\prime]}\mathcal{G}(X_i,A_i),
\end{equation} 
where $i^*$ represents the index of the optimal candidate.

In the fine-tuning phase, given a candidate setup $X=(x_1,\dots,x_L)^\intercal$, we randomly remove a device $x_l$ (with $1\leq l \leq L$). The setup $X$ is then updated if the QFI remains unchanged. It is worth noting that, within the main content of the article, we exhibit the original setups without any additional refinement in our results. However, this absence of modification bears no impact on the subsequent phase of response inference.

\subsection{Response Inferring}

From the recent work by \cite{alderete2022inference}, response function $R(\theta)$ can be precisely expressed as a trigonometric polynomial function. 
\begin{theorem}[Theorem 1, \cite{alderete2022inference}]
The defined response function $R(\theta)$ can be exactly expressed as a trigonometric polynomial of degree $n$:
\begin{equation}
    R(\theta)=\sum_{s=1}^n [a_s \cos(s\theta)+b_s \sin(s\theta)]+c,
\end{equation}
with $\{a_s,b_s\}_{s=1}^n$ and $c$ being real valued coefficients.
\end{theorem}
For trigonometric interpolation, the optimal approach is uniformly sampling the parameters as $\{\frac{2\pi (k-1)}{2n+1}\}_{k=1}^{2n+1}$.

The uncertainty or sensitivity of quantum sensing can be derived from error propagation $(\Delta\theta)^2=(\Delta R(\theta))^2/|\partial_\theta R(\theta)|^2$. This sensitivity is expressed in relation to the variance $(\Delta R(\theta))^2 = \mathrm{Tr}[S_\theta (\rho) O^2] - \mathrm{Tr}[S_\theta (\rho) O]^2$ and the slope of $R(\theta)$. When $O^2=I$, i.e. a Pauli string, this sensitivity can be represented as:
\begin{equation}
    (\Delta \theta)^2 = \frac{1-(\sum_{s=1}^n [a_s \cos(s\theta)+b_s \sin(s\theta)]+c)^2}{|\sum_{s=1}^n s[-a_s \sin(s\theta)+b_s \cos(s\theta)]|^2}.
\end{equation}

Moreover, for approximation $\tilde{R}(\theta)$ and the corresponding sensitivity $\Delta \tilde{\theta}$,  the estimation of sensitivity is related to the following theorem.
\begin{theorem}[Theorem 4, \cite{alderete2022inference}]
     Let $R(\theta)$ be the exact response function, and $\tilde{R}(\theta)$ be its approximation obtained from $M$-shot average $\overline{R}(\theta_k)$ with uniformly sample $\theta_k$. Defining the maximum estimation error $\epsilon = \mathrm{max}_{\theta_k \in P}|R(\theta_k)-\overline{R}(\theta_k)|$, and the slope of $R(\theta$ at a field $\theta_k$ as $D_l=|\partial_\theta \tilde{R}(\theta)|_{\theta=\theta_l}|$, then 
    \begin{equation}
        |\Delta \theta-\Delta\tilde{\theta}|\in \mathcal{O}(\frac{\epsilon \mathrm{log}(n)}{D_l}).
    \end{equation}
\end{theorem}
This indicates $\Delta\theta$ experiences fluctuations, especially in regions where the inferred response function's slope approaches zero.

\section{Extended results on eight-photon probe search}

Here we provide the discovered setup sequences in the eight-photon probe search task in the main text. The top-3 candidates identified in 50000 test examples are as follows
\begin{enumerate}
	\item $\text{PBS}_{b,c}\rightarrow\text{PBS}_{a,g}\rightarrow\text{QWP}_{h,0.5\pi}\rightarrow\text{PBS}_{d,f}\rightarrow\text{PBS}_{c,h}\rightarrow\text{R}_{d}\rightarrow\text{PBS}_{e,f}\rightarrow\text{HWP}_{a,0.5\pi}$; 
	\item $\text{PBS}_{a,g}\rightarrow\text{R}_{c}\rightarrow\text{PBS}_{b,c}\rightarrow\text{PBS}_{a,g}\rightarrow\text{PBS}_{g,f}\rightarrow\text{HWP}_{g,0.5\pi}\rightarrow\text{HWP}_{d,0.5\pi}\rightarrow\text{HWP}_{c,\pi}\rightarrow\text{R}_{b}\rightarrow\text{HWP}_{h,\pi}\rightarrow\text{QWP}_{a,\pi} $; 
	\item $\text{PBS}_{f,d}\rightarrow\text{R}_{a}\rightarrow\text{PBS}_{a,e}\rightarrow\text{QWP}_{b,0.25\pi}\rightarrow\text{PBS}_{h,a}$. 
\end{enumerate}

The top-1 candidate in 30000 training examples is
\begin{enumerate}
	\item $\text{R}_{b}\rightarrow\text{PBS}_{f,h}\rightarrow\text{QWP}_{h,0.75\pi}\rightarrow\text{QWP}_{f,\pi}\rightarrow\text{PBS}_{d,h}\rightarrow\text{QWP}_{e,0.5\pi}\rightarrow\text{R}_{c}\rightarrow\text{PBS}_{c,f}\rightarrow\text{R}_{f}\rightarrow\text{PBS}_{b,g}\rightarrow\text{QWP}_{g,0.75\pi}\rightarrow\text{R}_{c}\rightarrow\text{HWP}_{h,5\pi}$. 
\end{enumerate}

Here, the subscripts indicate the device parameters and applied photon paths (from $a$ to $h$). Besides, the initial state prepared by SPDC  are Bell state.

\section{Parameter quantization}

\begin{figure*}[ht]
\centering
\includegraphics[width=0.98\textwidth]{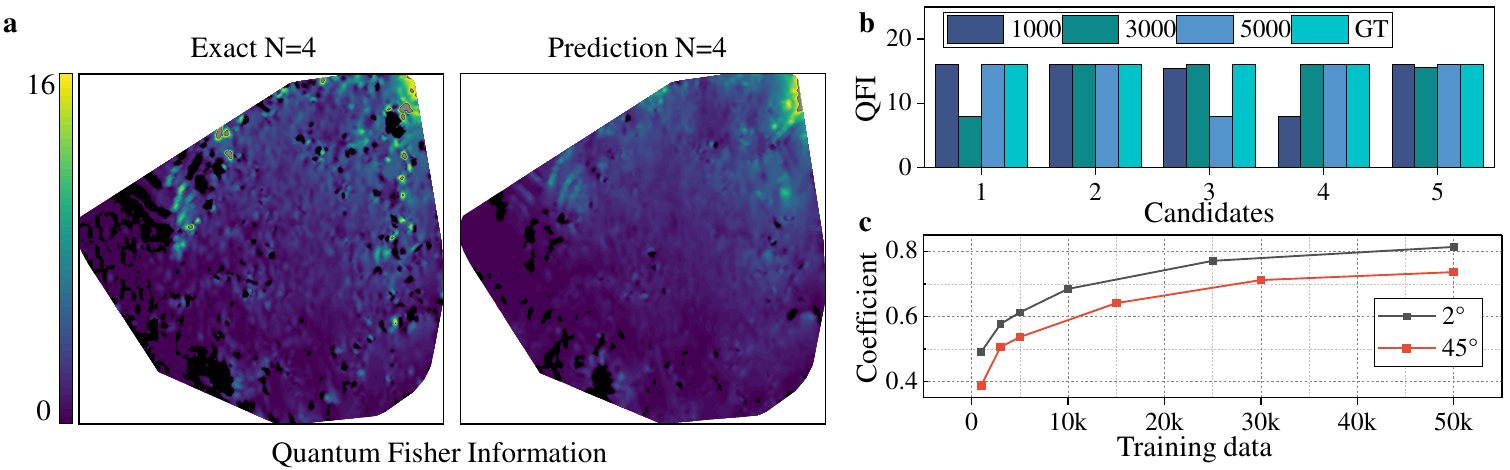}
\caption{\textbf{a. Latent Space Visualization.} On the left, exact QFIs for 4-photon optical setups are depicted, while on the right, the GNN predictions are shown. \textbf{b. Variation of Top-5 Candidates with Training Data Size.} The bars reflect the QFIs of the top-5 candidates from a pool of 10,000 test examples. These candidates are identified by a model trained on datasets of 1,000, 3,000, and 5,000 examples, juxtaposed against the ground truth. \textbf{c. Spearman's Correlation Coefficient between Predicted with Ground Truth QFIs.} The black and red lines represent the correlation coefficients for the $2^\circ$ and $45^\circ$ quantizations, respectively, and how they evolve with the volume of training data.}
\label{fig:sm2}
\end{figure*}

To transition a device with a continuous parameter into a one-hot vector representation, we discretize the possible angles of the wave-plate within the range $[0,\pi]$. As delineated in the main text, we've confined the angle to the set ${0,45^\circ, 90^\circ, 135^\circ}$. In this section, we delve into a more fine-grained quantization of the device parameters. Specifically, we quantize the angle in increments of $2^{\circ}$, encompassing potential angles of ${0,2^\circ,4^\circ,\dots,176^\circ,178^\circ}$.

For a comparative analysis, we employ the same 4-photon task with $H=\sum_{i=1}^4Z_i$. The latent space induced by the GNN is illustrated in Figure \ref{fig:sm2}a. It is evident that the transition from regions with lower QFI to those with higher QFI is more fluid in comparison to the results from the $45^\circ$ configuration. This can be attributed to the refined quantization, which engenders a smoother QFI distribution within the latent space.

Figures \ref{fig:sm2}b and c elucidate our model's performance relative to the size of the training dataset. A salient observation here is that the model exhibits superior performance under the fine-grained parameter setting. The Spearman correlation coefficient between the predicted and true QFI is notably higher compared to the results from the $45^\circ$ configuration. This might stem from the fact that a more diverse set of labels during training forces the model to master a robust predictor.

\section{Other encoding Hamiltonian}

\begin{figure*}[ht]
\centering
\includegraphics[width=0.98\textwidth]{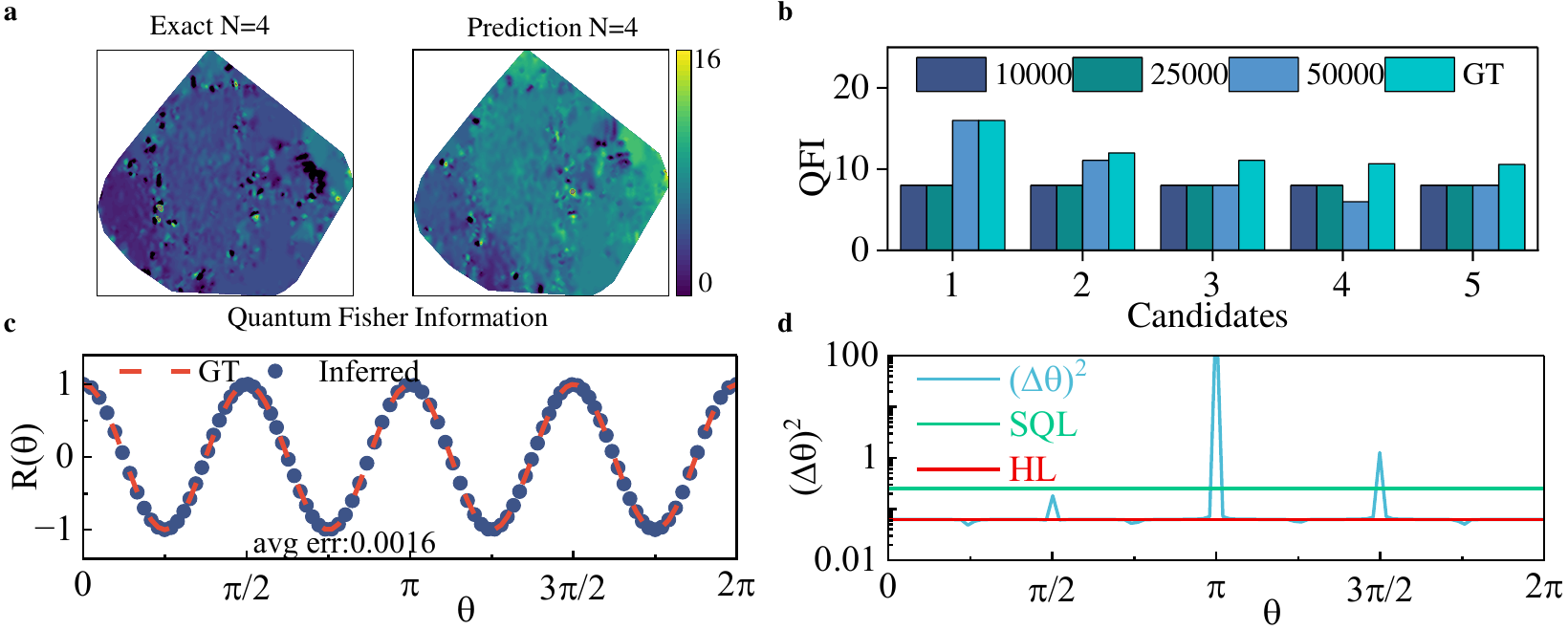}
\caption{\textbf{a. Latent Space Visualization.} On the left, exact QFIs for 4-photon optical setups are depicted, while on the right, the GNN predictions are shown. \textbf{b. Variation of Top-5 Candidates with Training Data Size.} The bars reflect the QFIs of the top-5 candidates from a pool of 10,000 test examples. These candidates are identified by a model trained on datasets of 1,0000, 25000, and 50000 examples, juxtaposed against the ground truth. \textbf{c. Inferred Response Functions.}
The inferred response function corresponds to the layout shown in Figure \ref{fig:sm4}\textbf{c}. The dashed red line represents the exact response function, while the points indicate the inferred response.
\textbf{d. Estimated Sensitivity.} The blue curve indicates the sensitivity level, while the green and red lines mark the SQL and HL, respectively. }
\label{fig:sm3}
\end{figure*}

The DQS does not depend on the explicit form of the encoding Hamiltonian, indicating its capability to discover the optimal probe state for various Hamiltonian. In this context, we consider a scenario where the parameter is encoded by the Hamiltonian $H =\sum_{i=1}^4 X_i$. 

Figures \ref{fig:sm3}a and b showcase the performance of the GNN in terms of QFI prediction and the search of the optimal probe state and optical setup. The prediction of QFI exhibits deviation from the exact value. Also, the number of training examples required is considerably higher compared to the case with $H =\sum_{i=1}^4 Z_i$. These findings suggest that the difficulty of seeking the optimal probe can differ based on the Hamiltonian.

Specifically, we consider 10000 test examples with only one optimal optical setup with QFI being 16. The initial 10,000 training examples lack any setup with this QFI. Upon adding another 15,000 examples, which include two optimal setups, the trained model still fails to identify the optimal test setup. We then continue to add 25000 training examples without optimal setups. This time, the trained model successfully discovers the optimal setup. These observations imply that the GNN is not simply memorizing the layouts of optimal setups but learning their pattern even from sub-optimal examples.

\begin{figure*}[ht]
\centering
\includegraphics[width=0.98\textwidth]{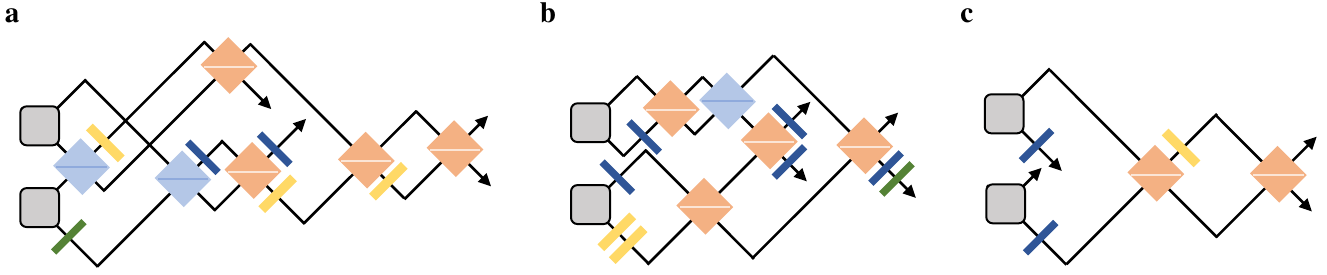}
\caption{\textbf{a-b} Illustration of optimal examples during the training phase. \textbf{c} Identified optimal setups during testing.}
\label{fig:sm4}
\end{figure*}

In Figure \ref{fig:sm4}, we detail the corresponding layouts of the training's optimal setups and the test's optimal one. Device sequences for the different setups are as follows:

\textbf{a}. $\text{QWP}_{d,0.25\pi} \rightarrow \text{PBS}_{b,c} \rightarrow \text{PBS}_{d,a} \rightarrow \text{R}_{d} \rightarrow \text{HWP}_{c,\pi} \rightarrow \text{BS}_{d,a} \rightarrow \text{HWP}_{d,0.25\pi} \rightarrow \text{BS}_{c,b} \rightarrow \text{BS}_{b,d} \rightarrow \text{R}_{a} \rightarrow \text{HWP}_{b,0.25\pi} \rightarrow \text{BS}_{d,b}$ with SPDC initial state $|\psi\rangle = (|0_a0_b\rangle+|1_a1_b\rangle)(|0_c0_d\rangle+|1_c1_d\rangle)$.

\textbf{b}. $\text{HWP}_{d,0.75\pi} \rightarrow \text{R}_{c} \rightarrow \text{HWP}_{d,0.5\pi} \rightarrow \text{BS}_{c,d} \rightarrow \text{R}_{b} \rightarrow \text{BS}_{b,a} \rightarrow \text{PBS}_{b,a} \rightarrow \text{BS}_{b,d} \rightarrow \text{BS}_{a,c} \rightarrow \text{R}_{a} \rightarrow \text{R}_{d} \rightarrow \text{QWP}_{a,0.75\pi} \rightarrow \text{R}_{b}$ with SPDC initial state $|\psi\rangle = (|0_a0_b\rangle+|1_a1_b\rangle)|1_c1_d\rangle$.

\textbf{c}. $\text{R}_{d} \rightarrow \text{BS}_{d,a} \rightarrow \text{HWP}_{d,0.25\pi} \rightarrow \text{R}_{b} \rightarrow \text{BS}_{a,d}$ with SPDC initial state $|\psi\rangle = (|0_a0_b\rangle+|1_a1_b\rangle)(|0_c0_d\rangle+|1_c1_d\rangle)$.

The probe states are $|\psi\rangle =\frac{1}{\sqrt{2}}(|++++\rangle + e^{-i\gamma}|----\rangle)$ with different phases $\gamma$ where $|+\rangle$ and $|-\rangle$ are the eigenvector of Pauli-X. The discovered optical setup in the test phase is notably more simple than setups seen during training, aligning with our previous analysis of the GNN's behavior.

For the quantum sensing task, we employ the measurement operator $O=\otimes_{i=1}^4 Z_i$ for the quantum sensing task. In Figure \ref{fig:sm4} c and d, we depict the inferred response function and the estimated sensitivity. The mean error is 0.0016, while the sensitivity achieves the Heisenberg limit.

\section{Two-photon interaction}

\begin{figure*}[ht]
\centering
\includegraphics[width=0.98\textwidth]{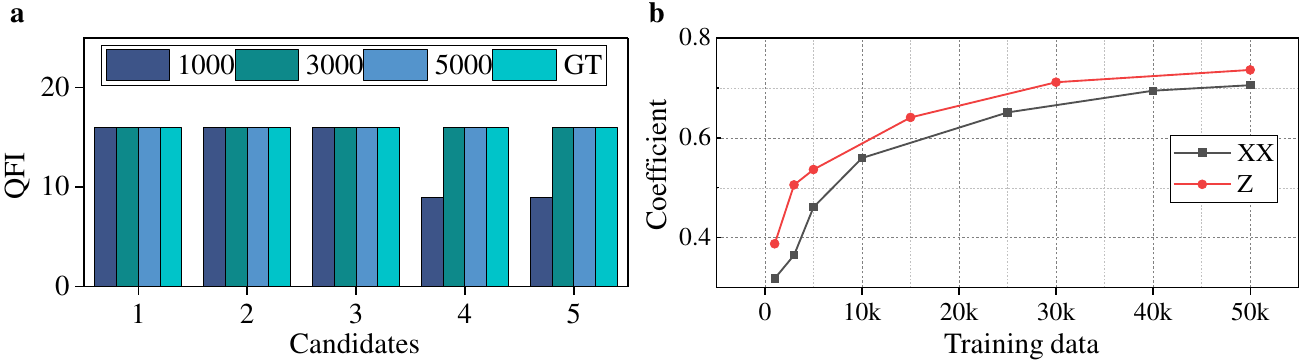}
\caption{\textbf{z}. Performance with respect to training data size. \textbf{b}. The Spearman coefficient between predicted QFIs and ground truth on 10000 test examples.}
\label{fig:sm5}
\end{figure*}

In this section, we delve into the encoding Hamiltonian that encompasses two-photon interactions, specifically represented as $H=\sum_{i<j}X_iX_j$. As depicted in Figure \ref{fig:sm5}, the GNN demonstrates proficiency in retrieving optimal setups even with a modest volume of training data. However, it's noteworthy that the Spearman correlation coefficient for this is a little lower in comparison to the scenario involving a one-photon interaction.

While certain instances have been reported where precision surpasses the Heisenberg Limit \cite{boixo2007generalized,napolitano2011interaction,yin2023experimental}, they do not directly align with the context of our discourse. Our study is specifically bounded to situations that consider a singular measurement operator.

\section{Computer-aided methods for optical quantum experiment}

In this section, we briefly overview existing computer-assisted algorithms for optical quantum experiments \cite{krenn2020computer}. 

The first is evolutionary algorithm  \cite{krenn2016automated,knott2016search,o2019hybrid,nichols2019designing}. The core idea is selecting the ``offspring'' experiment branching from a ``parent'' experiment. Their evaluation is done via a target function, keeping the satisfactory outcomes and discarding the unsatisfactory ones. For instance, in the work of Melvin \cite{krenn2016automated}, quantum experiments with randomized initialization are presented, leveraging symbolic algebra for the simulation and verification of quantum states. In a different work \cite{o2019hybrid}, deep neural networks is utilized for state classification, thereby eliminating the need for simulation and increasing the speed of the algorithm.

The second is the reinforcement learning algorithm \cite{hentschel2011efficient,lovett2013differential,melnikov2018active,xu2019generalizable,yang2020probe}, which is structured around the strategy of modifying actions as informed by a reward metric. For example in \cite{melnikov2018active}, the quantum experiment is treated as a series of actions conducted by an agent within an environment. This agent adjusts its actions, guided by active learning methods, according to the feedback it receives in the form of rewards.

The third is the gradient-based algorithm \cite{arrazola2019machine, krenn2021conceptual, ruiz2022digital}, which is similar to the variational quantum algorithm \cite{cerezo2021variational,zhou2023qaoa,tian2023recent}, where a quantum state is produced by parameterized optical quantum circuit, updated using gradient-based algorithms such as gradient descent. In the work of Pytheus \cite{krenn2021conceptual, ruiz2022digital}, they offer a unique perspective by representing a state through a parameterized graph. This graph, once optimized concerning a predefined objective function, is then converted into a quantum experiment.

One underlying thread connecting these methodologies is the pivotal role of feedback. Adjustments and modifications are contingent on this feedback, which can potentially lead to inefficiencies, especially during the vacuum period waiting for it. In practical applications, this bottleneck becomes even more evident when the exact form of the target is not pre-defined, making classical computer simulations unfeasible.

Recently, initial studies have leveraged deep learning models to characterize optical setups \cite{adler2021quantum,flam2022learning,jaouni2023deep} and learn quantum system \cite{torlai2018neural,carrasquilla2019reconstructing,palmieri2020experimental,ahmed2021quantum,gao2018experimental,yin2022efficient,koutny2023deep,zhu2022flexible,wu2023quantum}. These models learn quantum properties from collected offline data, rather than real-time interactions, thereby significantly increasing efficiency. Our work follows the paradigm of learning and exploits a more powerful deep-learning model for optical setups.

\end{document}